\documentclass[twocolumn,10pt]{asme2e}

\usepackage{lineno,hyperref}
\usepackage{subcaption}
\usepackage{graphicx}
\usepackage{lineno}
\usepackage{placeins}
\usepackage{array}
\usepackage{tabu}
\usepackage{bm}
\usepackage{amsmath}
\usepackage[dvipsnames]{xcolor}  
\usepackage{wasysym}
\usepackage{amssymb}
\modulolinenumbers[5]
\usepackage{algorithm, algorithmic}
\usepackage{mathtools, cuted}

\confshortname{}
\conffullname{}

\confdate{}
\confyear{}
\confcity{}
\confcountry{}

\papernum{}

\title{Learning Optimal Parametric Hydrodynamic Database for Vortex-Induced Crossflow Vibration Prediction}

\author{Samuel Rudy \\{\tensfb Jose del Aguila Ferrandis} \\{\tensfb Themistoklis Sapsis} \\{\tensfb Michael S. Triantafyllou}
    \affiliation{
	Department of Mechanical Engineering\\
	Massachusetts Institute of Technology\\
	Cambridge, MA 02139\\
    }	
}

\author{Dixia Fan 
    \affiliation{Department of Mechanical and Materials Engineering\\
    Queen’s University\\
    Kingston, Ontario K7M 3N9, Canada
    }
}

\begin{document}

\maketitle    
\begin{abstract}
{The Vortex-induced vibration (VIV) prediction of long flexible cylindrical structures relies on the accuracy of the hydrodynamic database constructed via rigid cylinder forced vibration experiments. However, to create a comprehensive hydrodynamic database with tens of input parameters including vibration amplitudes and frequencies and Reynolds number, surface roughness and so forth is technically challenging and virtually impossible due to the large number of experiments required. The current work presents an alternative approach to approximate the crossflow (CF) hydrodynamic coefficient database in a carefully chosen parameterized form. The learning of the parameters is posed as a constraint optimization, where the objective function is constructed based on the error between the experimental response and theoretical prediction assuming energy balance between fluid and structure. Such a method yields the optimal estimation of the CF parametric hydrodynamic database and produces the VIV response prediction based on the updated hydrodynamic database. The method then was tested on several experiments, including freely-mounted rigid cylinder in large Reynolds number with combined crossflow and inline vibrations and large-scale flexible cylinder test in the Norwegian Deepwater Program, and the result is shown to robustly and significantly reduce the error in predicting cylinder VIVs.}
\end{abstract}

\section*{INTRODUCTION}

The prediction of the bluff body vortex-induced vibrations (VIV) is a challenging but important problem. For example, if the VIV of a marine riser in the ocean current is not treated appropriately, it may result in severe fatigue damage and substantial economic and environmental loss. Therefore, due to its critical scientific and industrial applications, VIV has received a considerable amount of research attention in the last four decades. A large number of publications \cite{williamson2004vortex,gabbai2005overview,wang2020review} have reviewed the key concepts and principal mechanisms of the bluff body VIV's response and its wake patterns. 

One of the major difficulties in modeling and predicting the VIV of bluff body structures relies on the accuracy of the hydrodynamic force on the rigid cylinder and its distribution along the span of the flexible model \cite{fan2017vortex,FanVIVJFM2019,wang2021large}. The fluid force on oscillating cylinders varies significantly as a function of structural and flow properties \cite{williamson1996vortex, xu2013experimental,chen2013hydrodynamic}. Therefore, to quantitatively study the fluid forces on oscillating cylinders in the current, experiments were conducted, where a rigid cylinder is forced to vibrate with prescribed trajectories \cite{gopalkrishnan1993vortex,aronsen2007experimental,dahl2008vortex,fan2019robotic}. Especially for the CF vibration, the study focuses on the lift coefficient in-phase with the velocity $C_{lv}$, and the added mass coefficient in the CF direction $C_{my}$ as a function of true reduced frequency $f_r = \frac{fD}{U}$ and non-dimensional CF amplitude $A^* = \frac{A_y}{D}$, where $U$ is the prescribed fluid velocity, $f$ is the prescribed motion frequency, $A_y$ is the prescribed motion amplitude and $D$ is the cylinder diameter.

The measured hydrodynamic coefficients are not only helpful in understanding the nature of the rigid cylinder free vibration but also serve as the fluid force database of several semi-empirical codes \cite{triantafyllou1999pragmatic,roveri2001slenderex,larsen2001vivana} for flexible riser response in the ocean current, assuming that the strip theory is valid \cite{han2018hydrodynamic,FanVIVJFM2019,wang2021large}. However, to create a comprehensive hydrodynamic database with tens of input parameters including vibration amplitudes and frequencies as well as Reynolds number ($Re$) \cite{xu2013experimental}, surface roughness \cite{chang2011viv}, riser configurations \cite{lin2020dynamic} and so forth is technically challenging and virtually impossible due to the large number of experiments required. Besides, during the lifetime of a riser in the field, long-term effects, such as equipment aging and bio-fouling, inevitably alter the hydrodynamic coefficients, making long-term riser prediction and monitoring even more challenging \cite{meng2020fast}. 

Therefore, the current work presents an alternative approach to approximate the crossflow (CF) hydrodynamic coefficient database in a carefully chosen parameterized form. The learning of the parameters is posed as a constraint optimization, where the objective function is constructed based on the error between the experimental response and theoretical prediction assuming energy balance between fluid and structure. Such a method yields the optimal estimation of the CF parametric hydrodynamic database and produces the VIV response prediction based on the updated hydrodynamic database. The method then was tested on a freely-mounted CF-only vibrating rigid cylinder in a moderate $Re$ and a large-scale flexible cylinder test in the Norwegian Deepwater Program, and the result is shown to robustly and significantly reduce the error in predicting cylinder VIVs.

\section*{MATERIALS AND METHODS}

\begin{figure}
\centering
\includegraphics[width=\columnwidth]{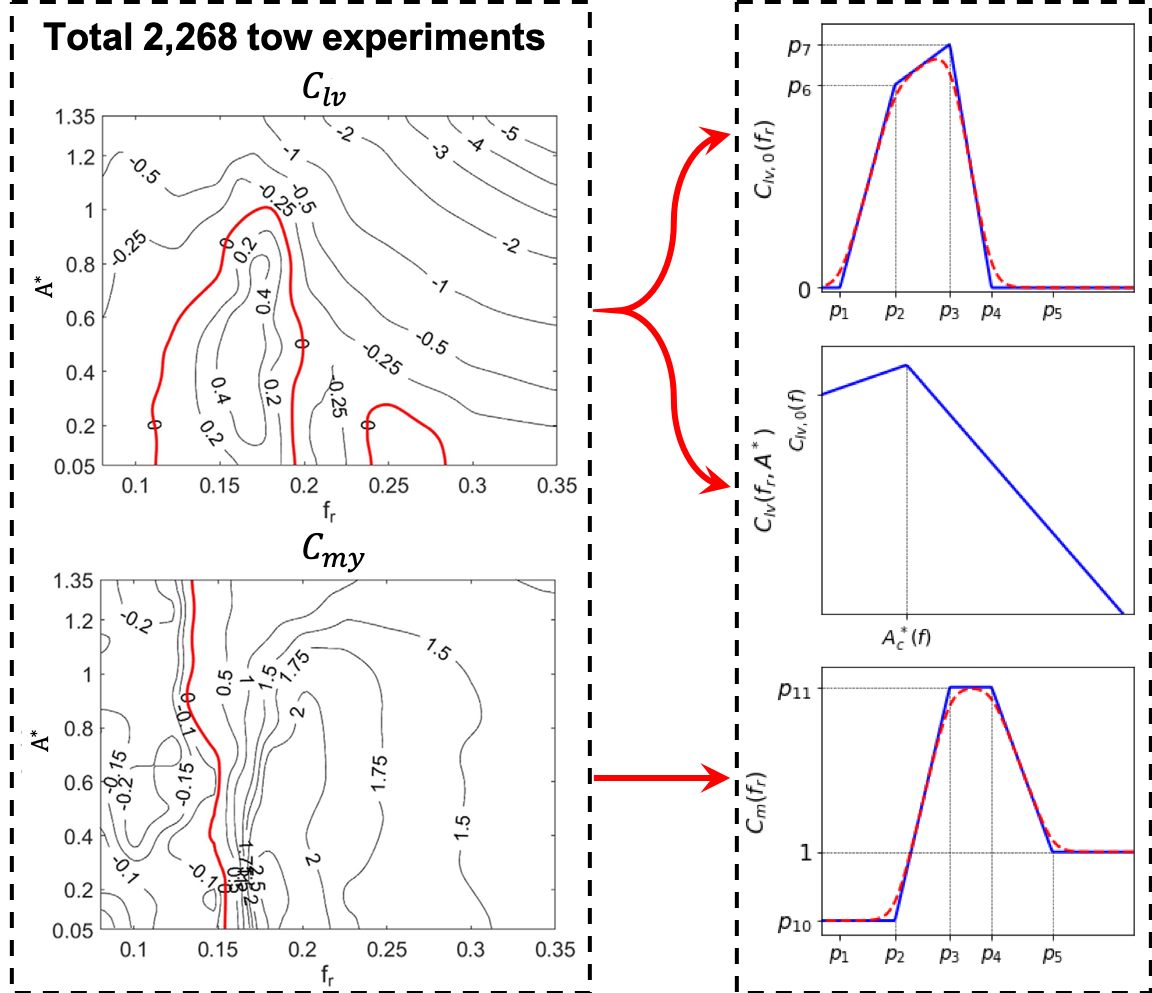}
\caption{The simplification of a hydrodynamic coefficient database constructed by a large number of rigid forced vibration experiments (left) \cite{fan2019robotic}, using simple piecewise linear models with a set of parameters to be learned from the data (right).}
\label{fig:piecewise_linear}
\end{figure}

\subsection*{Hydrodynamic Parameterization}

Shown in Fig. \ref{fig:piecewise_linear}, many of the desired properties of $C_{lv}$ and $C_{my}$ may be captured in a simple piecewise linear model: for example, the region of positive $C_{lv}$ is associate with particular $A^*$ and $f_r$ regions; $C_{my}$ is found to abruptly change from the small negative value to a large positive value at given $f_r$ and is weakly dependent on $A^*$. The above characters are given in a parameterized form quantified by parameters $p_i$ to be learned from the data.

In this work, we use a smoothed variation of the model shown in Fig. \ref{fig:piecewise_linear} to parameterize functional forms for the unknown hydrodynamic coefficients.  Rather than using a parameterization with sharp corners, we use softplus functions to parameterize each of the linear components of $C_m$, $C_{lv,0}$, and $(A_y/D)_c$ as this was found to improve fit.  $C_{lv}$ is left as a piecewise linear function so that it can be easily inverted to find $A^*$.  The exact functional forms for each quantity are given in the Appendix \ref{sec:exact_parameterization}.

\subsection*{Learning Parameters via Optimization}

We determine values of the parameters $\mathbf{p} = (p_1,\hdots,p_{11})$ through formulating and minimizing an appropriate error function been theoretical prediction using parameterized hydrodynamic database and given experimental observations. 

Take a rigid cylinder of mass $m$, diameter $D$ and length $L$, mounted on springs with spring constant $k$ and dashpots with damping constant $b$ in a uniform current of velocity $U$ as an example, its equation of motion is shown as follows, 

\begin{equation}
    m\frac{d^2y}{dt^2}+b\frac{dy}{dt}+ky = F_l(t),
    \label{eq:free}
\end{equation}
where $F_l(t)$ represents the oscillatory lift force acting from the fluid on the rigid cylinder. When the system achieves a harmonic oscillation assuming an energy balance between fluid and structure, we reach a set of equations for VIV response prediction in a non-dimensional form as follows,
\begin{equation}
\label{eq:CFprediction}
    V_r(U_r) = U_r \sqrt{\frac{m^*+C_{my}}{m^*+1}}, \quad A^*(U_r) = \frac{C_{lv}U^2_r}{4\pi^3(m^*+1)\zeta},
\end{equation}
where $V_r = \frac{U}{fD}$ is the true reduced velocity, the inverse of the true reduced frequency $f_r = \frac{fD}{U}$, $f$ is the response frequency, $U_r = \frac{U}{f_nD}$ is reduced velocity, $f_n = \frac{1}{2\pi}\sqrt{\frac{k}{m+\rho_f \forall}}$ is the system natural frequency in the still water, assuming $C_{my} = 1.0$, $m^* = \frac{\rho_s}{\rho_f}$ is the mass ratio, $\rho_s$ is the structural density, $\zeta$ is the damping ratio.

In the experiment, we normally do not have knowledge of the true hydrodynamic coefficients. Rather, we have measurements of $V_r$ and $A^*$ as a function of $U_{r}$ at a given $Re$. To determine the accuracy of any given parameterization we use Eq. \eqref{eq:CFprediction} and the parameterization given in the previous section to construct maps $V_r = V_r(U_r, \mathbf{p})$ and $A^* = A^*(U_r, \mathbf{p})$. This requires an inner loop solver to determine $U_{n}$ for arbitrary $V_r$. We minimize the residual of the first line of Eq. \eqref{eq:CFprediction} to find $V_r$.
\begin{equation}
    \hat{V}_r = \underset{V_r}{argmin} \left| V_r - U_r\sqrt{\frac{m^*+C_{my}(V_r;p)}{m^*+1}} \right|.
\end{equation}

The amplitude $A^*$ is subsequently found as the unique value which satisfies the second equation in Eq. \eqref{eq:CFprediction} given the previously found value of $V_r$ and parameterization $\mathbf{p}$. Comparing the observed and predicted reduced velocities and amplitudes, we get the learning objective function as follows, 
\begin{equation}
    J(\mathbf{p}) = \sum_{j=1}^n \frac{(f_{r,j} - \hat{f}_{r,j}(U_r, \mathbf{p}))^2}{\sigma^2_{f_r}} + \frac{(A^*_{j} - \hat{A}^*_{j}(U_r, \mathbf{p}))^2}{\sigma^2_{A^*}},
    \label{eq:free_coeff_error}
\end{equation}
where $\sigma^2_{f_r}$ and $\sigma^2_{A^*}$ are used for normalization.

The Eq. \eqref{eq:free_coeff_error} is appended with regularization terms and each $p_i$ is constrained to lie within a physically plausible region, described in the Appendix \ref{sec:exact_parameterization}. We solve for $\mathbf{p}$ using a randomized coordinate descent algorithm, which was found to perform favorably in comparison to gradient descent and particle swarm methods. Specifically, random perturbations of the current parameterization are taken along a specific axis, and the estimate is adjusted to the sample with the least error before moving to a new axis. The variance of the sampling distribution is reduced as optimization progresses. While heuristic, this algorithm was found to outperform all gradient-based methods, including gradient descent and BFGS, as well as particle swarm methods, simplex methods, and Bayesian optimization.

Similar to the rigid cylinder, the learning of hydrodynamic database of the flexible cylinder VIVs starts with the equation of the motion that describes a flexible cylinder along $z$-axis between $z = 0$ and $z = L$ with the mass per unit length $\mu$, the structural damping per unit length $c$, the bending stiffness of $EI$ taut and the tension $T$, under a time-varying fluid lift force per unit length $f_l(z, t)$ as follows, 
\begin{equation}
    \mu\frac{\partial^2y}{\partial t^2} + c\frac{\partial y}{\partial t} - \frac{\partial}{\partial z} (T \frac{\partial y}{\partial z}) + \frac{\partial^2}{\partial z^2}(EI\frac{\partial^2y}{\partial z^2}) = f_l(z,t).
    \label{eq:flexfree}
\end{equation}
By applying the assumption of the strip theory \cite{FanVIVJFM2019}, the free vibration response of a flexible cylinder in the current can be predicted with the forced vibration constructed hydrodynamic database. When the system achieves a harmonic oscillation, the problem can be modeled and solved as a nonlinear eigenvalue problem as follows,
\begin{equation}
\begin{split}
    [-\omega ^2(m+C_{my}\forall) + i\omega b]Y - \frac{\partial}{\partial z}(T\frac{\partial Y}{\partial z}) + \frac{\partial^2}{\partial z^2}(EI\frac{\partial^2Y}{\partial z^2}) \\= iC_{lv}\frac{\rho_f U^2}{2}D\frac{Y}{|Y|}, 
    \label{eq:egienFlex}
\end{split}
\end{equation}
where again, $C_{my}$ and $C_{lv}$ are functions of $A^*$ and $V_r$, which can be found in the hydrodynamic database constructed by the rigid cylinder forced vibration. From eq. \eqref{eq:egienFlex}, we are able to solve $\hat{V}_r(z,U_{r}^j(z),\mathbf{p})$ and $\hat{A}^*(z,U_{r}^j(z),\mathbf{p})$ along the span of the flexible cylinder via semi-empirical code VIVA \cite{triantafyllou1999pragmatic} using parameterized hydrodynamic database iteratively. Therefore, the objective function of the optimization can be imposed as follows, 
\begin{equation}
\begin{split}
    J(\mathbf{p}) = \sum_{j=1}^n (\frac{1}{L}\int|A^*_j - \hat{A}^*(z,U_{r}^j(z),\mathbf{p})|dz \\+ \lambda|V_r^j - \hat{V}_r(z,U_{r}^j(z),\mathbf{p})|,
    \label{eq:free_coeff_error2}
\end{split}
\end{equation}
where $\lambda$ is the weight between the model RMSE and the frequency error. Again the parameter $\mathbf{p}$ can be solved via the randomized coordinate descent algorithm used in the rigid cylinder problem. 

\section*{RESULT AND DISCUSSION}

In this section, we demonstrate the application of the proposed methodology to several datasets.  We start with the experimental datasets for CF-only rigid cylinder VIVs using force feedback apparatus. Then we show the application of the proposed method to large-scale flexible cylinder experiments in both uniform and linearly shear flow from the Norwegian Deepwater Programme.

\subsection*{Rigid Cylinders Free Vibration}

\begin{figure}
\centering
\includegraphics[width=\columnwidth]{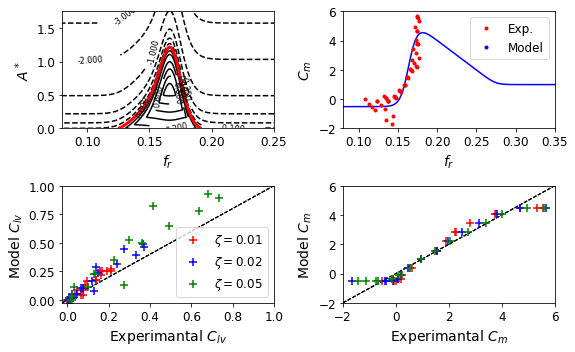}
\caption{Learned optimal hydrodynamic database of $C_{lv}$ (left column) and $C_{my}$ (right column). The second row shows the correlation between the experimental measurement and learned parametric model predicted hydrodynamic coefficients.}
\label{fig:predictfrAstar}
\end{figure}

Using a force feedback laboratory apparatus \cite{hover1997vortex}, Smogeli et al. \cite{smogeli2003force} studied CF-only free vibration of a rigid cylinder over a large range of $U_r$ with a fixed $Re = 20,000$. The structural properties were selected as a fixed mass ratio and various damping ratios. The constructed results of $C_{lv}$ and $C_{my}$ are plotted in Fig. \ref{fig:predictfrAstar} (top row). The result shows that learned $C_{lv}$ contour is similar to past CF-Only forced vibration result \cite{gopalkrishnan1993vortex} at a different $Re = 10,000$, and has a narrow positive region around $f_r = 0.16$ with maximum $A^*$ of zero contour line reach $A^* = 1.2$. Meanwhile, the learned $C_{my}$ is shown to have a drastic change from a small negative value to a large positive value at $f_r = 0.15$, similar to the findings in the past forced vibration experiments \cite{carberry2005controlled}.

In Smogeli et al.'s experiment, the hydrodynamic coefficients were measured. Therefore, we can compare the experimental measured hydrodynamic coefficient with our learned optimal $C_{lv}$ and $C_{my}$ database and plot the result in Fig. \ref{fig:predictfrAstar} (bottom row, right column). The two sets of data show good linear correlation indicating that the constructed hydrodynamic database using optimal parameters learned from the free vibration matches well with the experimentally measured vortex-induced force on the cylinder model. It is noted in Fig. \ref{fig:predictfrAstar} (bottom row, left column) that the predicted $C_{lv}$ is averagely larger than experimental measured $C_{lv}$, which may be a result of the experimental error due to the force-feedback setup \cite{hover1997vortex}. 

\begin{figure}
\centering
\includegraphics[width=\columnwidth]{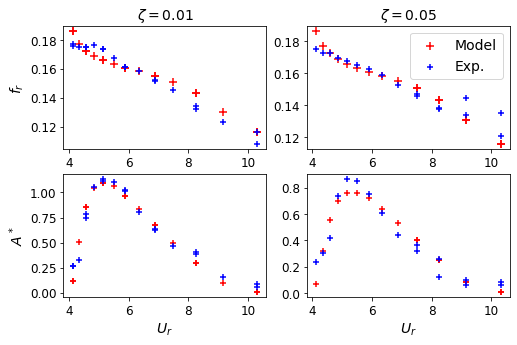} 
\caption{Comparison between experimental result (blue) and model prediction (red) of frequency (first row) and amplitude (second row) response for different damping ratios: $\zeta = 0.01$ (right column) and $\zeta = 0.05$ (left column).}
\label{fig:frAstar}
\end{figure}

With a good match between experiment measured hydrodynamic coefficient and learned fluid force database, the predicted structural response is found in good agreement with the experimental result for both displacement and frequency response, shown in Fig. \ref{fig:frAstar}. It is noteworthy that with an increase of $\zeta$, the maximum $A^*$ decreases. 

\subsection*{Flexible Cylinders Free Vibration}

We then test the learning method on the dataset of the Norwegian Deepwater Programme (NDP) large-scale laboratory experiment \cite{braaten2004ndp} of a 38m uniform flexible riser experiment in both the uniform current and linearly sheared flow with a wide range of different velocities.

\begin{figure}
\centering
\includegraphics[width=\columnwidth]{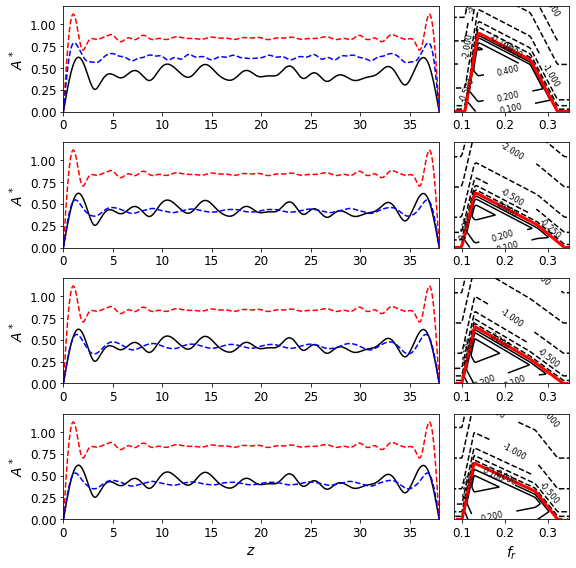}
\caption{Evolution of VIVA prediction on the NDP experiment no. 2182 (Uniform velocity $U = 2.0 m/s$) at iteration 0 (first row), iteration 2(second row), iteration 4(third row), iteration 6(fourth row). In the left column, the solid black line shows the experimental result; the red dashed line shows the nominal VIVA prediction, and the blue dashed line shows the VIVA prediction with corresponding learned hydrodynamic databases that are shown in the right column for $C_{lv}$. Additionally, the measured frequency response is 10.23 Hz in the experiment, the nominal VIVA prediction is 12.54Hz, and with the improved hydrodynamic database, the predicted frequency response evolves as 10.71 Hz (iteration 0), 10.62 (iteration 2), 10.56 (iteration 4) and 10.63 (iteration 6).}
\label{fig:sec0401}
\end{figure}

To demonstrate the learning process, we plot in Fig. \ref{fig:sec0401} the evolution of the VIVA prediction with the optimized hydrodynamic database over initial guess, first, second and sixth iterations of the $U = 2.0 m/s$ case (NDP experiment no. 2182). The left column shows the comparison between the prediction and the experiment. From the result, we can see with the increase of the iteration, the difference between the experiment and VIVA prediction with an optimized hydrodynamic database improves with the increase of the iteration. Furthermore, the last row of the displacement plot shows that with a learned optimal hydrodynamic database (solid blue line), VIVA can provide a better prediction than that using the standard hydrodynamic database (red dashed line), comparing to the experimental result (solid black line). The right column in Fig. \ref{fig:sec0401} plots the corresponding $C_{lv}$ of learned parameters $\mathbf{p}$, and it can be shown that the hydrodynamic database $C_{lv}$ does not vary much at larger iterations, representing a convergence of the learning process. Additionally, the measured frequency response is 10.23 Hz in the experiment, the nominal VIVA prediction is 12.54Hz, and with the improved hydrodynamic database, the predicted frequency response evolves as 10.71 Hz (the initial guess), 10.62 (the second iteration), 10.56 (the fourth iteration) and 10.63 (the sixth iteration). Moreover, we can conclude with the converged optimal hydrodynamic database; VIVA can provide a better prediction on the frequency response.

\begin{figure}
\centering
\includegraphics[width=\columnwidth]{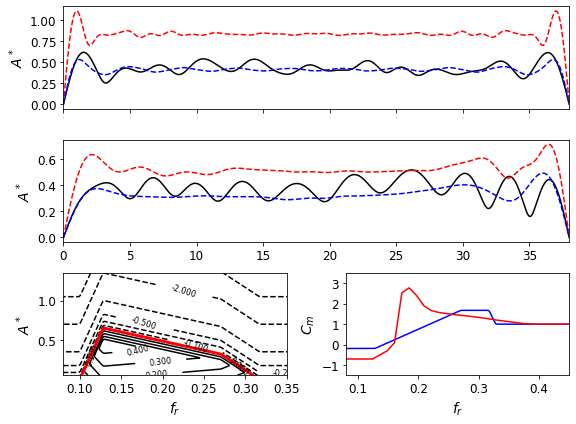}
\caption{The sample cases of no. 2182 cases of uniform flow (uniform $U = 2.0m/s$, first row) and no. 2430 cases of linearly sheared flow (maximum $U = 1.5m/s$, second row) are selected to show the difference of the amplitude response among the experimental result (solid black line), nominal VIVA prediction (red dashed line) and VIVA prediction with optimal hydrodynamic database (blue dashed line). The third row shows the learned hydrodynamic database of $C_{lv}$ (left) and $C_{my}$ (right, red line: $C_{my}$ for nominal VIVA and blue line: learned $C_{my}$ based on the NDP experiment).}
\label{fig:sec0402}
\end{figure}

Using a dataset of cases of uniform flow and cases of linearly sheared flow, we obtained the optimal hydrodynamic database of $C_{lv}$ and $C_{my}$, shown in the third row of Fig. \ref{fig:sec0402}. As an example, no. 2182 cases of uniform flow (uniform $U = 2.0m/s$) and no. 2430 cases of linearly sheared flow (maximum $U = 1.5m/s$) are selected out and plotted in the first and second row of Fig. \ref{fig:sec0402}. The result can be shown that with an optimally learned hydrodynamic database, the VIVA prediction (solid blue line) has been shown to match better with the experimental result (solid black line) than that of the VIVA prediction using a standard hydrodynamic database (red dashed line).

\begin{figure}
\centering
\includegraphics[width=\columnwidth]{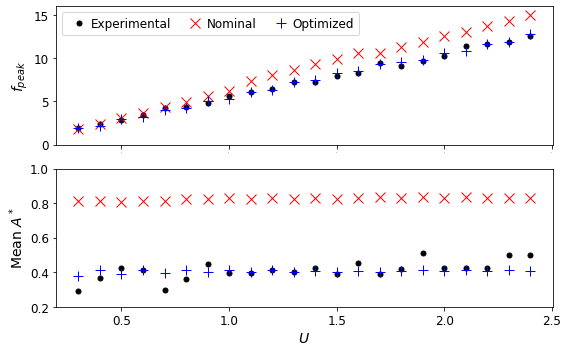}
\caption{Comparison of spatial average displacement $\overline{A^*}$ among experimental result (black dot), nominal VIVA prediction (red cross), and VIVA prediction with optimal hydrodynamic database for NDP 38m experiment in uniform flow over different incoming velocity $U$ in a unit of $m/s$.}
\label{fig:sec0403}
\end{figure}

\begin{figure}
\centering
\includegraphics[width=\columnwidth]{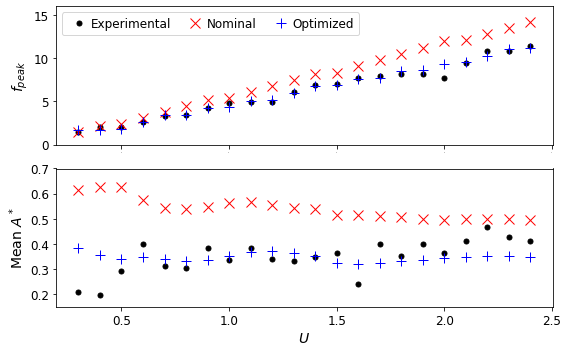}
\caption{Comparison of spatial average displacement $\overline{A^*}$ among experimental result (black dot), nominal VIVA prediction (red cross), and VIVA prediction with optimal hydrodynamic database for NDP 38m experiment in linearly sheared flow over different maximum incoming velocity $U$ in a unit of $m/s$.}
\label{fig:sec0404}
\end{figure}

In order to summarize the improvements made by the optimally learned hydrodynamic database to the standard one.  We plot the peak response frequency $f_{peak}$ and the mean disturbance $\overline{A^*} = \frac{1}{L}\int_{0}^{L}A^*dz$ for all velocity cases of the NDP experiment for both uniform (in Fig. \ref{fig:sec0403}) and linearly sheared flow (in Fig. \ref{fig:sec0404}) cases. The result clearly shows that the prediction with the optimized hydrodynamic database (blue cross) exhibits greater accuracy in both frequency and amplitude prediction than the prediction with the standard database (red star) and that improvements are universal across all velocities.

\section*{CONCLUSION}
In this work, we present a methodology for determining parametric forms of hydrodynamic databases that can be trained with sparse sensors along a riser's length. The parametric structure of the proposed databases is informed by experiments performed on rigid cylinders forced vibrations. The corresponding parameters can be learned via optimization from the dataset of both rigid and flexible cylinders VIVs.  We show that the proposed method yields more accurate predictions for VIV amplitude and frequency both in rigid and flexible cylinder experiments.


\bibliographystyle{asmems4}

\begin{acknowledgment}
We gratefully acknowledge support from the DigiMaR Consortium, consisting of ABS, ExxonMobil, Petrobras, SAIPEM, Shell, and Subsea 7. 
\end{acknowledgment}

\bibliography{asme2e}

\begin{thebibliography}{10}

\bibitem{williamson2004vortex}
Williamson, C., and Govardhan, R., 2004.
\newblock ``Vortex-induced vibrations''.
\newblock {\em Annu. Rev. Fluid Mech., \textbf{ 36}}, pp.~413--455.

\bibitem{gabbai2005overview}
Gabbai, R., and Benaroya, H., 2005.
\newblock ``An overview of modeling and experiments of vortex-induced vibration
  of circular cylinders''.
\newblock {\em Journal of Sound and Vibration, \textbf{ 282}}(3-5),
  pp.~575--616.

\bibitem{wang2020review}
Wang, J., Fan, D., and Lin, K., 2020.
\newblock ``A review on flow-induced vibration of offshore circular
  cylinders''.
\newblock {\em Journal of Hydrodynamics, \textbf{ 32}}(3), pp.~415--440.

\bibitem{fan2017vortex}
Fan, D., and Triantafyllou, M.~S., 2017.
\newblock ``Vortex induced vibration of riser with low span to diameter ratio
  buoyancy modules''.
\newblock In The 27th International Ocean and Polar Engineering Conference,
  International Society of Offshore and Polar Engineers.

\bibitem{FanVIVJFM2019}
Fan, D., Wang, Z., Triantafyllou, M., et~al., 2019.
\newblock ``Mapping the properties of the vortex-induced vibrations of flexible
  cylinders in uniform oncoming flow''.
\newblock {\em Journal of Fluid Mechanics}, p.~to be appear.

\bibitem{wang2021large}
Wang, Z., Fan, D., Triantafyllou, M.~S., and Karniadakis, G.~E., 2021.
\newblock ``A large-eddy simulation study on the similarity between free
  vibrations of a flexible cylinder and forced vibrations of a rigid
  cylinder''.
\newblock {\em Journal of Fluids and Structures, \textbf{ 101}}, p.~103223.

\bibitem{williamson1996vortex}
Williamson, C.~H., 1996.
\newblock ``Vortex dynamics in the cylinder wake''.
\newblock {\em Annual review of fluid mechanics, \textbf{ 28}}(1),
  pp.~477--539.

\bibitem{xu2013experimental}
Xu, Y., Fu, S., Chen, Y., et~al., 2013.
\newblock ``Experimental investigation on vortex induced forces of oscillating
  cylinder at high reynolds number''.
\newblock {\em Ocean Systems Engineering, \textbf{ 3}}(3), pp.~167--180.

\bibitem{chen2013hydrodynamic}
Chen, Y., Fu, S., Xu, Y., et~al., 2013.
\newblock ``Hydrodynamic characters of a near-wall circular cylinder
  oscillating in cross flow direction in steady current''.
\newblock {\em Acta Physica Sinica, \textbf{ 62}}(6), p.~064701.

\bibitem{gopalkrishnan1993vortex}
Gopalkrishnan, R., 1993.
\newblock ``Vortex-induced forces on oscillating bluff cylinders''.
\newblock PhD thesis, Massachusetts Institute of Technology.

\bibitem{aronsen2007experimental}
Aronsen, K.~H., 2007.
\newblock ``An experimental investigation of in-line and combined in-line and
  cross-flow vortex induced vibrations''.
\newblock PhD thesis, Norwegian University of Science and Technology.

\bibitem{dahl2008vortex}
Dahl, J.~M., 2008.
\newblock ``Vortex-induced vibration of a circular cylinder with combined
  in-line and cross-flow motion''.
\newblock PhD thesis, Massachusetts Institute of Technology.

\bibitem{fan2019robotic}
Fan, D., Jodin, G., Consi, T., et~al., 2019.
\newblock ``A robotic intelligent towing tank for learning complex
  fluid-structure dynamics''.
\newblock {\em Science Robotics, \textbf{ 4}}(36).

\bibitem{triantafyllou1999pragmatic}
Triantafyllou, M., Triantafyllou, G., Tein, Y., et~al., 1999.
\newblock ``Pragmatic riser viv analysis''.
\newblock In Offshore technology conference, Offshore Technology Conference.

\bibitem{roveri2001slenderex}
Roveri, F.~E., and Vandiver, J.~K., 2001.
\newblock ``Using shear7 for assessment of fatigue damage caused by current
  induced vibrations''.
\newblock In Proc. 20th OMAE Conf, pp.~3--8.

\bibitem{larsen2001vivana}
Larsen, C.~M., Vikestad, K., Yttervik, R., et~al., 2001.
\newblock ``Vivana theory manual''.
\newblock {\em Marintek, Trondheim, Norway}.

\bibitem{han2018hydrodynamic}
Han, Q., Ma, Y., Xu, W., et~al., 2018.
\newblock ``Hydrodynamic characteristics of an inclined slender flexible
  cylinder subjected to vortex-induced vibration''.
\newblock {\em International Journal of Mechanical Sciences, \textbf{ 148}},
  pp.~352--365.

\bibitem{chang2011viv}
Chang, C. C.~J., Kumar, R.~A., and Bernitsas, M.~M., 2011.
\newblock ``Viv and galloping of single circular cylinder with surface
  roughness at 3.0$\times$ 104< re <1.2$\times$ 105''.
\newblock {\em Ocean Engineering, \textbf{ 38}}(16), pp.~1713--1732.

\bibitem{lin2020dynamic}
Lin, K., Fan, D., and Wang, J., 2020.
\newblock ``Dynamic response and hydrodynamic coefficients of a cylinder
  oscillating in crossflow with an upstream wake interference''.
\newblock {\em Ocean Engineering, \textbf{ 209}}, p.~107520.

\bibitem{meng2020fast}
Meng, X., Wang, Z., Fan, D., Triantafyllou, M., and Karniadakis, G.~E., 2020.
\newblock ``A fast multi-fidelity method with uncertainty quantification for
  complex data correlations: Application to vortex-induced vibrations of marine
  risers''.
\newblock {\em arXiv preprint arXiv:2012.13481}.

\bibitem{hover1997vortex}
Hover, F., Miller, S., and Triantafyllou, M., 1997.
\newblock ``Vortex-induced vibration of marine cables: experiments using force
  feedback''.
\newblock {\em Journal of fluids and structures, \textbf{ 11}}(3),
  pp.~307--326.

\bibitem{smogeli2003force}
Smogeli, O.~N., Hover, F.~S., and Triantafyllou, M.~S., 2003.
\newblock ``Force-feedback control in viv experiments''.
\newblock In International Conference on Offshore Mechanics and Arctic
  Engineering, Vol.~36835, pp.~685--695.

\bibitem{carberry2005controlled}
Carberry, J., Sheridan, J., and Rockwell, D., 2005.
\newblock ``Controlled oscillations of a cylinder: forces and wake modes''.
\newblock {\em J. Fluid Mech., \textbf{ 538}}, pp.~31--69.

\bibitem{braaten2004ndp}
Braaten, H., and Lie, H., 2004.
\newblock ``{NDP} riser high mode {VIV} tests''.
\newblock {\em Norwegian Marine Technology Research Institute, Technical
  Report}(512394.00), p.~01.

\end{thebibliography}

\appendix
\section{Parametric Hydrodynamic Databases}
\label{sec:exact_parameterization}

\begin{figure}
\centering
\includegraphics[width=\columnwidth]{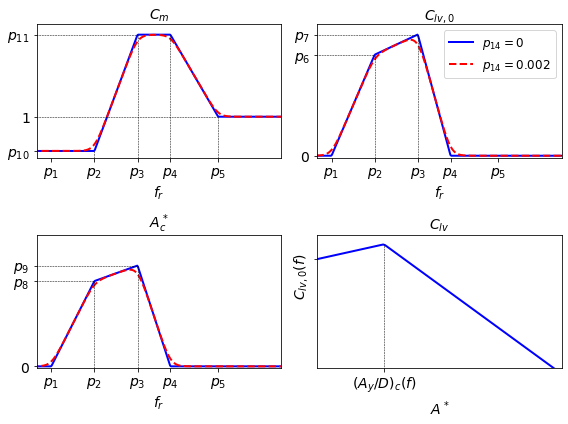}
\caption{parameters}
\label{fig:appa01}
\end{figure}

The hydrodynamic coefficients are defined as functions of reduced frequency and amplitude using a fourteen dimensional parametric representation.  The effect of the first thirteen of these parameters is shown in Fig. \ref{fig:appa01} with the fourteenth parameter acting as a smoothing factor for transitions between linear components.  Descriptions of each parameter are given in table \ref{tab:p_description}.  Note that in the non-smoothed case the parameters $p_6-p_9$ define the exact values of $C_{lv,0}$ and $A^*_c$ at frequencies $p_2$ and $p_3$ but when we use a smoothed representation this will no longer be the case.

\begin{table}[!ht]
\centering
\begin{tabular}{| c | c |} 
\hline
 $p_1-p_5$ & $f_r$ used for $C_m$, $C_{lv,0}$, and $A^*_c$.\\
 $p_6,p_7$ & Parameters for values of $C_{lv,0}$ at $p_2$, $p_3$.  \\
 $p_8,p_9$ & Parameters for values of $A^*_c$ at $p_2$, $p_3$.  \\
 $p_{10},p_{11}$ & Values of constant sections of $C_m$ curve.  \\
 $p_{12}$ & Slope of $C_{lv}$ w.r.t. $A^*$ for $A^*<A^*_c$  \\
 $p_{13}$ & Slope of $C_{lv}$ w.r.t. $A^*$ for $A^*>A^*_c$  \\
 $p_{14}$ & Characteristic width of softplus function.  \\ [1ex] 
 \hline
\end{tabular}
\caption{Description and bounds of parameters for hydrodynamic coefficients}
\label{tab:p_description}
\end{table}

The smoothed forms of $C_m$, $C_{lv,0}$, and $A^*_c$ are given in terms of the soft-plus function which is defined by,
\begin{equation}
    s_w(f) = w\, \log (1+exp(f/w))
    \label{eq:softplus}
\end{equation}
where $w$ is a learned length scale.  As $w\to 0$ the soft-plus function converges uniformly to the rectified linear unit, allowing for smooth approximation of piece wise linear functions.  The added mass coefficient is given by,
\begin{equation}
\begin{aligned}
    C_m &= p_{10} + m_1 (s_{p_{14}}(f-f_2) - s_{p_{14}}(f-f_3)) \\
    &+ m_2 (s_{p_{14}}(f-f_4) - s_{p_{14}}(f-f_5))
\end{aligned}
\label{eq:Cm_softplus}
\end{equation}
where,
\begin{equation}
    m_1 = \frac{p_{11}-p_{10}}{p_3-p_2}
    \hspace{1 cm} \text{and} \hspace{1 cm}
    m_2 = \frac{1-p_{11}}{p_5-p_4}.
\end{equation}
The values of $C_{lv,0}$ and $A^*_c$ are given by,
\begin{equation}
\begin{aligned}
    C_{lv,0} &= m_3(s_{p_{14}}(f-f_1) - s_{p_{14}}(f-f_2)) + m_4(s_{p_{14}}(f-f_2) \\ &- s_{p_{14}}(f-f_3)) + m_5(s_w(f-f_3) - s_{p_{14}}(f-f_4))
\end{aligned}
\label{eq:Clv0_softplus}
\end{equation}
and
\begin{equation}
\begin{aligned}
    A^*_c &= m_6(s_{p_{14}}(f-f_1) - s_{p_{14}}(f-f_2)) + m_7(s_{p_{14}}(f-f_2) \\ &- s_{p_{14}}(f-f_3)) + m_8(s_{p_{14}}(f-f_3) - s_{p_{14}}(f-f_4)),
\end{aligned}
\label{eq:ADc_softplus}
\end{equation}
where
\begin{equation}
    \begin{aligned}
    &m_3 = \frac{p_{6}}{p_2-p_1},\hspace{1 cm}m_4 = \frac{p_{7}-p_{6}}{p_3-p_2}, \hspace{1 cm}m_5 = \frac{-p_{7}}{p_4-p_3}\\
    &m_6 = \frac{p_{8}}{p_2-p_1}, \hspace{1 cm} m_7 = \frac{p_{9}-p_{8}}{p_3-p_2}, \hspace{1 cm} m_8 = \frac{-p_{9}}{p_4-p_3}
    \end{aligned}
\end{equation}
Finally, $C_{lv}$ is given by,
\begin{equation}
    C_{lv} = \left\{ \begin{aligned}
    &C_{lv,0} + A^* p_{12}, \text{  if  } A^* \leq A^*_c\\
    & C_{lv} = C_{lv,0} + A^*_c p_{12} - (A^* - A^*_c)p_{13}, \text{  if  } A^* > A^*_c
    \end{aligned}\right. 
\end{equation}
Each variable is bounded based on the intervals described in Tab. \ref{tab:p_bounds}.  Open boundaries are enforced through fitting the transformed variables $q_i = \sigma^{-1}((p_i-p_{min})/(p_{max}-p_{min}))$ where $p_{min}$ and$p_{max}$ are the bounds for the given parameter and $\sigma$ is the standard sigmoid function.
\begin{table}[!ht]
\centering
\begin{tabular}{| c | c |} 
\hline
 $p_1$ & $(0.08,0.245)$\\
 $p_n$, $n=2,3,4,5$ & $(p_{n-1},0.35)$\\
 $p_6-p_7$ & $(0,0.25)$\\
 $p_8-p_9$ & $(0,2)$\\
 $p_{10}$ & $(-2,1) / (-2,0)$\\
 $p_{11}$ & $(1,10) / (1,5)$\\
 $p_{12}$ & $(0.1,5)$\\
 $p_{13}$ & $(1,5)$\\
 $p_{14}$ & $(0,0.005)$\\
 \hline
\end{tabular}
\caption{Description and bounds of parameters for hydrodynamic coefficients.  Case with two sets of values indicate rigid / flexible VIV databases.}
\label{tab:p_bounds}
\end{table}


\end{document}